\documentclass{optica-article}

\journal{opticajournal} 

\articletype{Research Article}

\usepackage{xspace}
\usepackage{amsmath}
\usepackage{graphicx}
\usepackage{color}

\newcommand{\etal}{\textit{et al.\@}\xspace}
\newcommand{\invitro}{\textit{in vitro}\xspace}

\newcommand{\enface}{\textit{en face}\xspace}

\newcommand{\enfaceh}{\textit{en-face}\xspace}
\newcommand{\um}{\(\muup\)m\xspace}
\newcommand{\uM}{\(\muup\)M\xspace}
\newcommand{\dbsq}{dB$^2$\xspace}

\graphicspath{{./figure}{.}}

\begin{document}

\title{Neural-network-based high-speed and high-definition full-field dynamic optical coherence tomography}

\author{Suzuyo Komeda,\authormark{1}
		Nobuhisa Tateno,\authormark{1}
		Yusong Liu,\authormark{1}
		Rion Morishita,\authormark{1}
		Xibo Wang,\authormark{1}
		Ibrahim Abd El-Sadek,\authormark{1,2}
		Atsuko Furukawa,\authormark{3}
		Satoshi Matsusaka,\authormark{3}
		Shuichi Makita,\authormark{1} 
		and Yoshiaki Yasuno\authormark{1,*}}

\address{\authormark{1}Computational Optics Group, University of Tsukuba, Tsukuba, Ibaraki, Japan.\\
		\authormark{2}Department of Physics, Faculty of Science, Damietta University, New Damietta City, Damietta, Egypt.\\
		\authormark{3}Clinical Research and Regional Innovation, Faculty of Medicine, University of Tsukuba, Tsukuba, Ibaraki, Japan.
		}

\email{\authormark{*}yoshiaki.yasuno@cog-labs.org}

\begin{abstract*}
A neural-network (NN)-based method for high-speed, high-definition dynamic optical coherence tomography (DOCT) using full-field swept-source optical coherence microscopy (FF-SS-OCM) is demonstrated.
FF-SS-OCM provides high-definition OCT images, but, particularly in DOCT imaging, it results in a significant enlargement of the data size and subsequently long data streaming and processing time, which prevents high-throughput imaging.
We address this issue by introducing an NN-based DOCT method that generates high-definition logarithmic intensity variance (LIV) -based DOCT images from only four OCT volumes, whereas the conventional method required 32 volumes.
The NN model successfully generates an LIV image that is qualitatively and quantitatively similar to the LIV image computed from 32 volumes.
This approach significantly reduces data size, transfer time, and processing time for DOCT imaging by a factor of eight.
Specifically, these were reduced from 42 GB to 5.3 GB, 7 min to 55 s, and 4 hours to 30 min, respectively. 
\end{abstract*}

\section{Introduction}
In recent years, advancements in cultivation technology have made \invitro samples increasingly functional\cite{Knight2014Anatomy} and allowed them to better mimic living tissues\cite{Caddeo2017Frontier}. 
Consequently, \invitro samples have become essential for drug development and large-scale screening.
Simultaneously, the advancement of \invitro samples has increased the need for high-quality imaging modalities to extract their structural and functional information.

To achieve the structural imaging, bright-field microscopy has been typically used due to its high resolution\cite{Fei2022Bioeng}.
However, since it lacks depth resolution, the structural imaging cannot be achieved volumetrically.
In contrast, to extract the functional information, fluorescence microscopy has been widely used\cite{Lichtman2005NatMes, Kessel2017SLAS, Lee2018Theran}.
Although fluorescence microscopy has target cell specificity, the procedure usually requires markers. 
And hence, this modality is typically invasive.
Additionally, the imaging penetration is limited to within a-few-hundred micrometers.

Optical coherence tomography (OCT) is an emerging modality that provides non-invasive and volumetric imaging.
It has been used for the imaging of \invitro samples, such as tumor spheroids\cite{Huang2017CanRes}, and has visualized their internal structure. 

However, two problems make OCT a sub-optimal modality. 
One is insensitivity to intratissue activity. 
The other is a trade-off between lateral resolution and depth of focus (DOF). 
The first problem has been recently addressed by dynamic OCT (DOCT), which visualizes intratissue activity without any labels\cite{Azzollini2023BOE, Ren2024ComBio, Josefsberg2025BOE, Heldt2025BOE}.
The second problem has been partially resolved by several computational refocusing methods, such as digital holographic methods that invert defocusing\cite{Yasuno2006OE, LYu2007OE} and interferometric synthetic aperture microscopy\cite{Ralston2006OSA, Ralston2007NatPhy, YZLiu2017BOE}.
However, there is an inevitable problem for widely used conventional point-scanning OCT.
Since the conventional point-scanning OCT has a confocal pinhole, it reduces the singly scattered light from a scatterer as the defocus increases.
Namely, the confocal pinhole still limits the imaging depth even if the lateral resolution is restored by the computational refocusing\cite{Zhu2025arXiv}.
To overcome this point, full-field swept-source optical coherence microscopy (FF-SS-OCM) has been recently developed\cite{Hillmann2016SciRep, Hillmann2017OE, Borycki2019BOE, Tateno2025arXiv}.
Due to the lack of a confocal pinhole, FF-SS-OCM is not affected by depth-dependent signal decay caused by the confocality.
Additionally, it enables the usage of a high-NA objective, which leads to high lateral resolution.
By combining FF-SS-OCM with computational refocusing, high lateral resolution and long DOF are achieved simultaneously\cite{Hillmann2016SciRep, Tateno2025arXiv}.

Recently, FF-SS-OCM has been combined with DOCT to make OCT much more suitable for \invitro sample imaging.
This combination, denoted as FF-DOCT, can achieve high-resolution and label-free intratissue activity imaging with a longer imaging depth than conventional point-scanning OCT\cite{Tateno2025arXiv}.

However, FF-DOCT still suffers from a significant problem caused by its high-resolution nature.
Since high-resolution imaging naturally requires high pixel densities, i.e., high definition, the data size of FF-SS-OCM becomes inevitably large.
In addition, DOCT requires a large number of repeatedly acquired OCT images at the same location.
Consequently, the combination of FF-SS-OCM with DOCT further increases the data size, which leads to significantly long data streaming time and long data processing time.
For instance, the FF-SS-OCM developed in Ref.\@ \cite{Tateno2025arXiv} provides 896 $\times$ 768 pixels in the lateral direction and requires a 32-image repeating acquisition for DOCT imaging, which leads to around 10 min including measurement time and data transfer time from the camera to the PC. 
For example, if all samples in a 96-well plate need to be measured, data acquisition takes over 10 hours.
Additionally, data processing time, including OCT reconstruction, computational refocusing, and DOCT computation, takes around 4 hours even for a single DOCT dataset. 
This problem significantly hinders the application of FF-DOCT for high-throughput imaging such as large-scale screening.

To reduce the number of images for DOCT, neural networks (NNs) have been recently employed. 
Liu \etal demonstrated NN-based DOCT generation\cite{Liu2024BOE}, which can generate a logarithmic intensity variance (LIV) image \cite{ElSadek2020BOE, ElSadek2021BOE} from only 4 images, which is one-eighth of the conventionally required 32 images.
However, the application of this method to FF-DOCT is not straightforward.
First, the original NN model in Ref.\@ \cite{Liu2024BOE} was solely trained by the conventional point-scanning OCT data, and hence this ready-made NN model cannot be used directly with FF-SS-OCM data.
Another strategy could be training an NN model from scratch using FF-SS-OCM data.
However, this strategy is unfeasible because it requires a large number of FF-DOCT datasets and, as mentioned, the acquisition and computation of a single FF-DOCT volume take more than 4 hours.
To overcome these issues, we hypothesized that a model refinement strategy, that is additional training with a limited amount of FF-DOCT data on the NN model established for point-scanning OCT, would make the model applicable to FF-SS-OCM data and enable high-speed FF-DOCT imaging.

In this paper, we demonstrate high-speed FF-DOCT generation, where the NN model established for point-scanning OCT with relatively low resolution \cite{Liu2024BOE} (denoted as the basement NN model) is additionally trained by a limited number of FF-DOCT datasets to generate an LIV image from only 4 OCT images.
The refined NN-generated LIV image was subjectively and objectively compared with conventional LIV images (i.e., the ground truth images) and also with the basement NN-generated LIV images.
Consequently, the size of an FF-DOCT dataset is reduced from 42 GB to 5.3 GB.
This data size reduction shortened the data transfer time and the data processing time from around 7 min to 55 s and 4 hours to 30 min, respectively.

\section{Principle}
\subsection{Basement NN model}
\label{sec:basementModel}
\begin{figure}
	\centering\includegraphics[width=14cm]{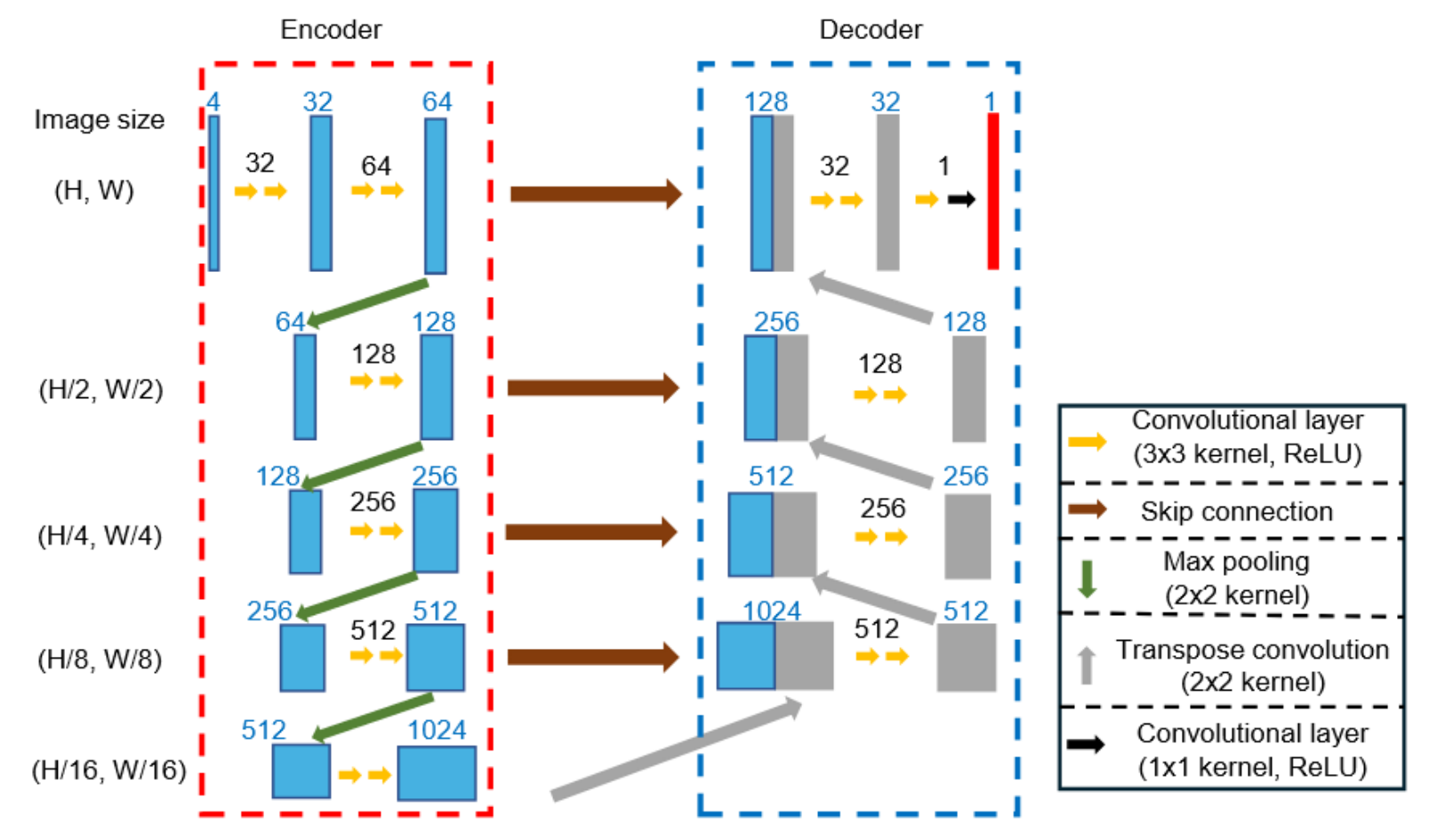}
	\caption{%
		NN architecture used for LIV generation.
		The NN consists of three parts: an encoder (red dashed-line box), a decoder (blue dashed-line box), and a skip connection (brown arrows).
		The input to the NN is a set of four cross-sectional OCT images, which were acquired at the same position in the sample but at different time points, and the output is an LIV image, which is a DOCT image defined as the time variance of the dB-scaled OCT images.
		The leftmost part indicates the spatial dimensions of the image at each step of the U-Net, where (H, W) represents the original size of the input image.
	}
	\label{fig:modelArchi}
\end{figure}
Since the refinement strategy is fully based on the basement NN model, its brief explanation is given here.
The basement NN model is identical to the one established in our previous study for a point-scanning swept-source OCT system with a relatively low resolution and a probe wavelength of 1.3 \um \cite{Liu2024BOE}.
The model architecture is a modified version of the U-Net architecture\cite{Ronneberger2015Springer}, as shown in Fig.\@ \ref{fig:modelArchi} (that is identical to the architecture shown in Ref.\@ \cite{Liu2024BOE}).
The input has four channels corresponding to four time-sequential cross-sectional dB-scaled OCT images, and the output is a DOCT image.

The target DOCT image used as a ground truth is an LIV image, which is defined as the time variance of a dB-scaled OCT image computed from 32 time-sequential OCT images\cite{ElSadek2021BOE} as follows
\begin{equation}\label{eq:LIV}
	\mathrm{LIV}(x, z) = \left\langle \left( \mathrm{I}_\text{dB}(x, z, t_i) - \langle \mathrm{I}_\text{dB}(x, z, t_i) \rangle_i \right)^2 \right\rangle_i,
\end{equation}
where $\text{I}_\text{dB}(x, z, t_i)$ represents a dB-scaled OCT intensity image at the spatial position $(x, z)$, and $x$ and $z$ are the lateral and depth positions, respectively.
$t_i$ is the time point of acquisition of the $i$-th frame. 
$\langle \; \rangle_i$ represents the average over time points.
LIV is mainly sensitive to the occupancy of dynamic scatterers among all scatterers in the resolution volume\cite{Morishita2025BOE}.

In our typical implementation, the OCT time sequence used to compute the ground truth LIV consists of 32 images, and the time separation between consecutive images (i.e., $t_{i+1} - t_i$) is 204.8 ms, which means that the entire time sequence is acquired in 6.35 s.
The detailed implementation of this time-sequential data acquisition (i.e., the scan protocol) is described elsewhere \cite{ElSadek2021BOE}.

For the NN input, the image sequence consists of four images with a time separation between adjacent images of 1,638.4 ms, which is eight times longer than that of the ground truth LIV image.
The total time separation from the first frame to the last is 4.92 s.
The four dB-scaled OCT images are then concatenated into a four-channel dataset to be fed into the NN model.
Specifically for the training process, the 8th, 16th, 24th, and 32nd images of a 32-image sequence are extracted to make the input image sequence, while the LIV computed using all the 32 images is used as a ground truth (i.e., the target).

To establish the basement NN model, a large number of datasets was used in the training process.
Specifically, 57,600 image patches generated from 72 samples were used for training, and 19,200 image patches generated from 24 samples were used for validation. 
Here each patch is a small region (64 $\times$ 64 pixels) of an image.
In total, 76,800 image patches generated from 96 samples were used. 
The samples used for the basement model training included 48 human breast cancer (MCF-7 cell line) spheroids and 48 colon cancer (HT-29 cell line) spheroids.
The spheroids were treated with anti-cancer drugs, which were paclitaxel for MCF-7 spheroids and SN-38 for HT-29 spheroids with concentrations of 0, 0.1, 1, and 10 \uM over treatment durations of 1, 3, and 6 days.

The system used for the data acquisition was a swept-source point-scanning OCT\cite{ELi2017BOE, Miyazawa2019BOE}.
The center wavelength is 1.3 \um.
The resolutions are 19 \um at $1/e^2$-width for the lateral direction and 19 \um in air (14 \um in tissue) at full-width at half-maximum (FWHM) for the axial direction.
The lateral pixel separations were 1.95 \um and 7.81 \um respectively for the fast- and slow-scan directions, and the axial pixel separation was 7.24 \um.
The field of view (FOV) was 1 mm $\times$ 1 mm.

In the training process, weighted mean absolute error (wMAE) was used as a loss function.
Here wMAE was selected rather than conventional MAE because the LIV values in the ground truth LIV image were not evenly distributed, i.e., low LIV values were dominant.
The weight was customized to increase the weights of the high-LIV regions.
Specifically, LIV values of 9 \dbsq or more were weighted by a factor of 2. 
More details are given in Eqs. (2) and (3) in Ref.\@ \cite{Liu2024BOE}.

\subsection{Model refinement}
\label{sec:modelRefine}
\begin{figure}
	\centering\includegraphics[width=14cm]{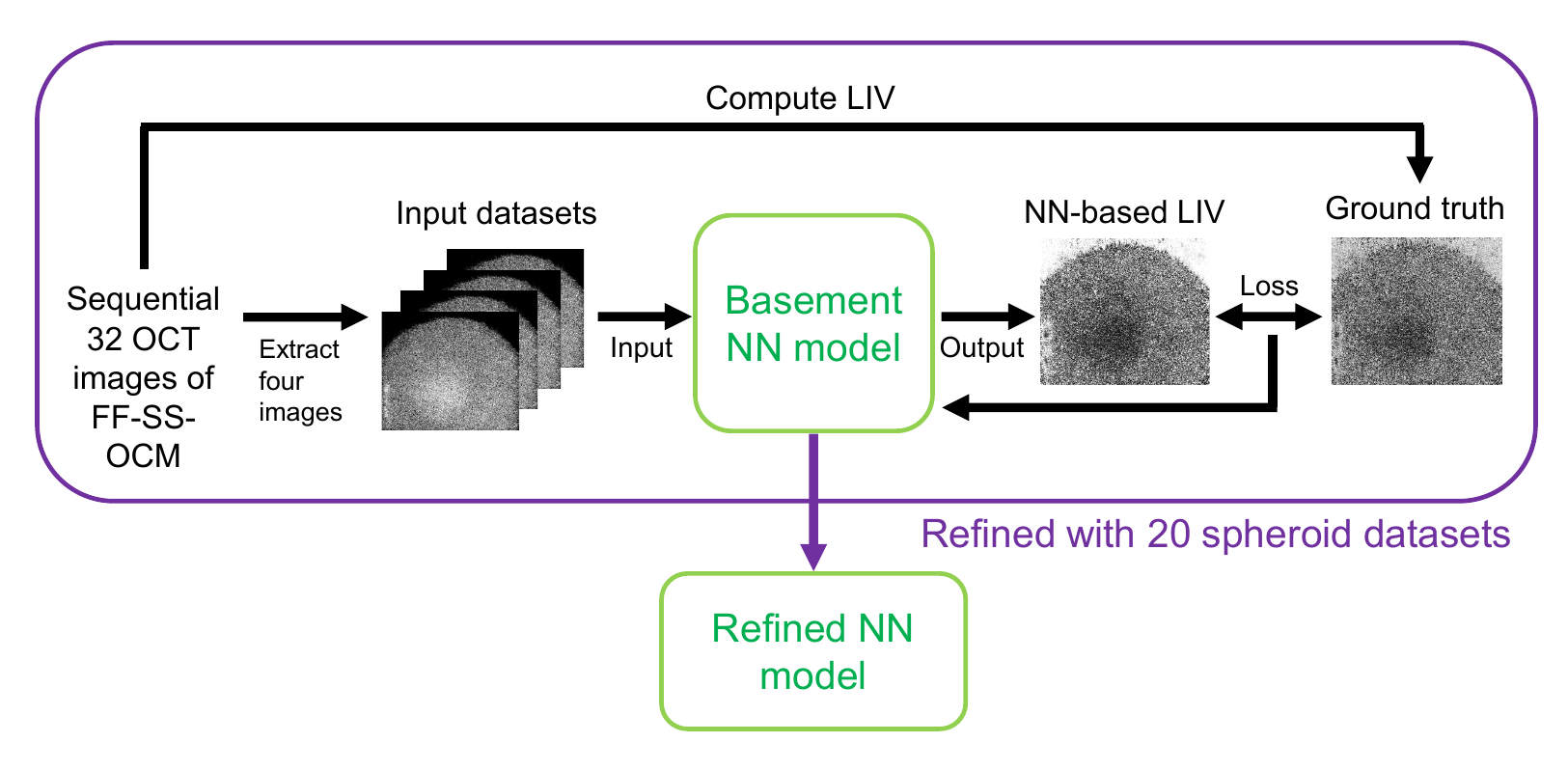}
	\caption{%
		Schematic diagram of the model refinement.
		The basement model has been trained by low-resolution point-scanning OCT images.
		Refinement is performed with FF-SS-OCM data sequences of only 20 spheroids, and finally gives the refined NN model.
	}
	\label{fig:refineFlow}
\end{figure}
Thus far, we have briefly described the basement NN model.
To make the NN model applicable to FF-SS-OCM data, namely, to enable the model to generate an LIV image from only 4 images of FF-SS-OCM data, refinement is performed on the basement model using a small number of FF-SS-OCM data as shown in Fig.\@ \ref{fig:refineFlow}.

A single FF-DOCT dataset consisted of 32 time-sequential \enfaceh OCT images.
From these 32 images, four (8th, 16th, 24th, and 32nd) OCT images are extracted and fed into the basement NN model.
The output image is compared with the ground truth LIV computed from the 32 OCT images, and the loss is computed.
The model is then updated using this loss.
This process is repeated with the small number of FF-SS-OCM datasets.
Finally, we obtain the refined NN model, which can generate an LIV image by inputting four time-sequential logarithmic OCT images of the FF-SS-OCM data.
The detailed implementation of the refinement process is described in Section \ref{sec:refineImplement}.

\section{Implementation}
\label{sec:implementation}

\subsection{OCT system and DOCT acquisition protocol of FF-SS-OCM}
\label{sec:DOCTacqui}
\begin{figure}
	\centering\includegraphics[width=14cm]{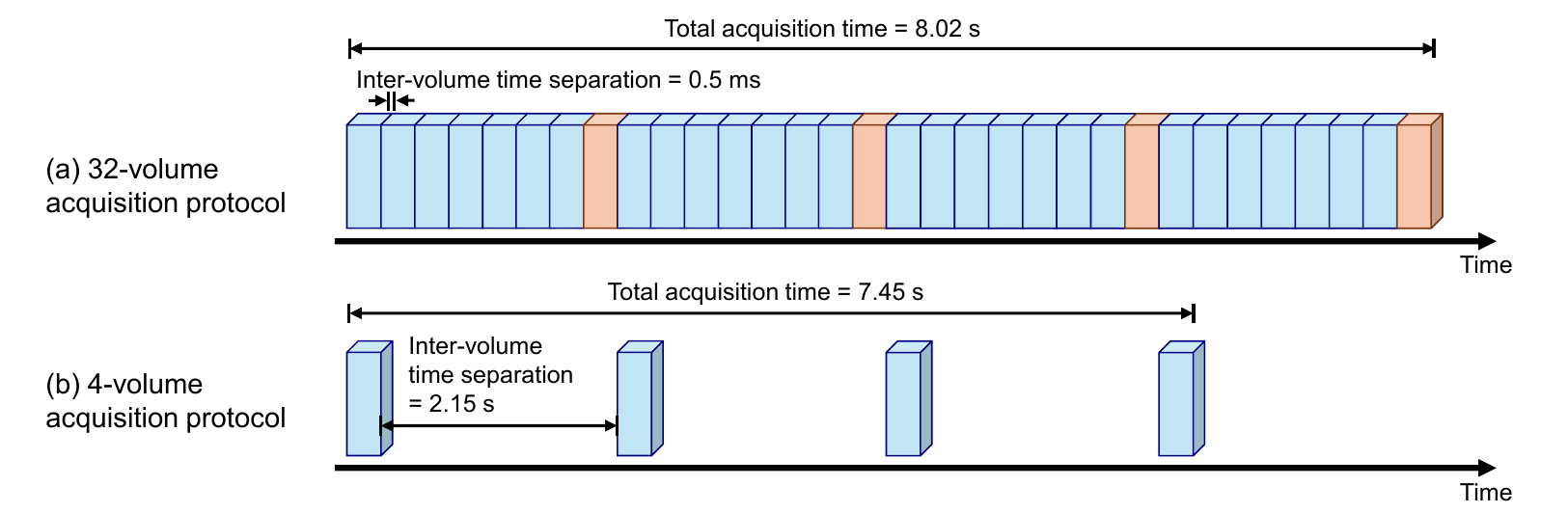}
	\caption{%
		Schematic diagram of two types of volume acquisition protocols.
		(a) 32-volume acquisition protocol, where the inter-volume time separation is 0.5 ms and the total acquisition time is 8.02 s.
		The four volumes with orange color are extracted from the 32 volumes as input to the NN.
		(b) Four-volume acquisition protocol, where the time separation is 2.15 s and the total acquisition time is 7.45 s.
		The inter-volume time separation of 2.15 s is identical to that of the four volumes extracted from the 32 volumes.
		}
	\label{fig:acquiProto}
\end{figure}
A custom-built FF-SS-OCM was used for data acquisition.
Since full details of this system can be found elsewhere \cite{Tateno2025arXiv}, here we only briefly describe the system specifications.
The system employs a sweeping light source with a center wavelength of 840 nm (BS-840-1-HP, Superlum, Ireland).
The interference signals of the sample and reference lights are captured by a high-speed two-dimensional CMOS camera (maximum frame rate = 6,400 frames/s for 1024 $\times$ 1024 pixels, FASTCAM Mini AX100, Photron, Japan).
Because this system uses high-NA objectives (NA = 0.3, RMS10X-PF, Olympus, Japan), it provides 1.4-\um lateral resolution at the FWHM of the intensity.
The axial resolution is 6.5 \um in air and 4.7 \um in tissue.

We use 32-volume acquisition protocol for DOCT imaging as depicted in Fig.\@ \ref{fig:acquiProto}(a).
For a single volume acquisition, 1,000 \enfaceh spectral interferograms are obtained at 4,000 frames/s.
And hence, one volume is acquired in 0.25 s.
The inter-volume time separation between the last spectral interferogram of a volume and the first interferogram of the next volume is 0.5 ms.
Hence, the total acquisition time for the 32 volumes becomes 8.02 s (see also Fig.\@ \ref{fig:acquiProto} for the terminology definitions).
The transversal field of view (FOV) is 0.43 mm $\times$ 0.37 mm with 896 $\times$ 768 pixels.
The pixel separations are 0.48 \um $\times$ 0.48 \um in the lateral direction and 3.41 \um in the axial direction.

\subsection{DOCT reconstruction}
\label{sec:DOCTreconstruction}
The raw spectral interference signals are first rescaled into the $k$-domain and inverse-Fourier transformed to obtain complex OCT signals.
Then an \enfaceh spatial frequency filter is applied to the complex OCT to remove artifacts in the high-frequency region (i.e., outside the cut-off frequency of the imaging optics).
Finally, computational refocusing is performed by applying a phase-only spatial frequency filter\cite{Makita2025BOE}.
Note that the identical spatial-frequency filter and the phase-only refocusing filter are used over the 32 volumes.
Finally, LIV is computed as the logarithmic variance of the dB-scaled OCT intensity as shown in Eq.\@ (\ref{eq:LIV}).
More details of the signal processing are published elsewhere \cite{Tateno2025arXiv}.

\subsection{Samples for model refinement}
\begin{table}
	\caption{%
		List of ten cultivation and treatment conditions of spheroids used in the refinement process.
		Twenty spheroid individuals were used.
		``Total'' column shows the total number of spheroid at each condition.
		The ``Training'' and ``Validation'' columns indicate the number of spheroids used for the training and validation, respectively. 
		The 20 spheroids were split into 16 and 4 for training and validation, respectively.
		Although the number of the spheroids in each condition was not evenly distributed, it was ensured that the training datasets included all the sample types.
	}
	\centering
	\footnotesize
	\begin{tabular}{cccccc}
		\hline
		Cultivation period {[}day{]} & DOX concentration {[}\uM{]} & Treatment period {[}day{]} & Total & Training & Validation \\
		\hline
		9  & 0  & 0 & 5  & 4  & 1 \\
		6  & 0  & 1 & 2  & 2  &   \\
		6  & 1  & 1 & 2  & 1  & 1 \\
		6  & 10 & 1 & 1  & 1  &   \\
		8  & 0  & 3 & 2  & 1  & 1 \\
		8  & 1  & 3 & 2  & 2  &   \\
		8  & 10 & 3 & 1  & 1  &   \\
		11 & 0  & 6 & 2  & 1  & 1 \\
		11 & 1  & 6 & 1  & 1  &   \\
		11 & 10 & 6 & 2  & 2  &   \\
		\hline
		&   & Total  & 20 & 16 & 4         
	\end{tabular}
	\label{tab:dataSplitting}
\end{table}

In the refinement process, datasets obtained from 20 breast cancer (MCF-7 cell line) spheroids were used.
Each spheroid was formed by initially seeding 1,000 tumor cells and cultivated in each well of a U-shaped 96-well plate.
For each measurement, one spheroid was extracted from the well plate and placed on a Petri dish.
Among the 20 spheroids, 5 spheroids were cultivated for 9 days without any drugs applied.
The remaining 15 spheroids were treated with an anti-cancer drug (doxorubicin; DOX) on day 5 of cultivation with concentrations of 0, 1, and 10 \uM over treatment durations of 1, 3, and 6 days.
In total, 10 types of spheroids (i.e., three drug concentrations times three treatment durations, and non-treated) were used.

The 20 spheroids were divided into 16 for training and 4 for validation.
The specific data splitting of each spheroid type is detailed in Tab.\@ \ref{tab:dataSplitting}.
Although the number of the spheroids in each condition was not evenly distributed, it was ensured that the training datasets included all the sample types.

\subsection{Implementation of refinement}
\label{sec:refineImplement}
\subsubsection{Datasets for model refinement}
To generate the training and validation datasets, the spheroids were measured by the 32-volume sequential acquisition protocol [Fig.\@ \ref{fig:acquiProto}(a)].
For each DOCT dataset, the ground truth LIV was computed from 32 volumes, and a corresponding input four-volume sequence was generated by extracting four volumes from 32 volumes.
The series of four volumes has an inter-volume time separation of 2.15 s.

From each input dataset, 20 \enfaceh images containing at least 6,000 pixels in the sample region were extracted.
Note that for point-scanning OCT, the inputs are cross-sectional images, while here the input images are \enface because, for FF-SS-OCM, image observation by \enfaceh slices is primary.
Then, from each \enfaceh image, 40 patch pairs with 64 $\times$ 64 pixels were selected.
Therefore, both for training and validation, the sizes of the input image and the ground truth LIV image were 64 $\times$ 64 pixels.
The extracted location for each patch was selected at random, but all the patches were enforced to contain the spheroid region at least partially.
The patch locations could also partially overlap.
The final training and validation datasets consisted of 12,800 and 3,200 patch pairs, respectively.

\subsubsection{Detailed implementation of refinement}
\label{sec:detailedImplementaion}
The refinement was performed on the basement NN model as keeping all the hyperparameters previously used in its training process\cite{Liu2024BOE} except for the loss function.
In the refinement training, the 12,800 patch pairs were grouped into 800 mini batches, where each mini batch consisted of 72 image patch pairs.

Similar to the case of point-scanning OCT, the ground truth LIV values obtained by FF-SS-OCM were not evenly distributed.
And hence, we used weighted mean absolute error (wMAE) as a loss function.
We customized the weight $w$ to increase the weights of the high-LIV regions as follows
\begin{equation} \label{eq:weight}
	W(x, y, b; \theta) =
	\begin{cases}
		w & \text{for } \text{LIV}(x, y; b) \ge \theta \\
		1 & \text{otherwise}
	\end{cases},
\end{equation}
where $w > 1$, LIV is the ground truth LIV image defined by Eq.\@ (\ref{eq:LIV}), $(x, y)$ is a spatial position within the batch, $b$ is the batch index, and $\theta$ is a predefined threshold for the LIV. 
As shown in this equation, we set high weights for the pixels whose ground truth LIV values are larger than the threshold value ($\theta$). 
In this study, we set $w$ and $\theta$ at 3 and 30 \dbsq, respectively.

The wMAE was then computed from the NN output and the ground truth using this weight as follows
\begin{equation}
	\text{wMAE}(\theta) = \frac{\sum_{x, y, b} \left[W(x, y, b; \theta) \circ \left|\text{LIV}(x, y; b) - \text{LIV}'(x, y; b)\right|\right]}{\sum_{x, y, b} W(x, y, b; \theta)},
\end{equation}
where $\text{LIV}$ and $\text{LIV}'$ represent the ground truth LIV and the NN output, respectively, and $\circ$ represents the element-wise product.
The rationality of this loss function is discussed later in Section \ref{sec:optimalRefine}.

The refined NN model parameters were updated using the Adam optimizer\cite{Kingma2014arXiv} with the wMAE loss.
To enable the NN model to learn detailed image patterns, we used a decaying learning rate strategy\cite{You2019arXiv}, where the learning rate was defined as $10^{-4} + 5 \times 10^{-4}/\text{epoch}$.
To prevent over-fitting, the training process in refinement was stopped when the validation loss did not decrease for seven consecutive epochs, and the parameters at which the validation loss was the smallest were stored as the refined NN model.

The model refinement was implemented in Python 3.8 using the open-source machine learning platform TensorFlow 2.3.0. 
It was performed on a PC equipped with a graphics processing unit (GPU; NVIDIA GeForce RTX 3070 Ti with 6,144 Compute Unified Device Architecture (CUDA) cores and 64 GB of memory).

\subsection{Protocol and method of performance evaluation study}
\label{sec:evaluationStudy}
We used three types of images for the performance evaluation of the refined NN model.
The first image type is a standard LIV image computed from 32 dB-scaled OCT volumes, and this LIV image was identical to the ground truth image.
This type is denoted as ``ground truth (GT).''
The second image type is the output image of the refined NN model, and this image type is denoted as ``refined LIV.''
The last image type is the output image of the basement NN model, namely, non-refined NN model, and this image type is denoted as ``non-refined LIV.''

To enable subjective observation of the image, pseudo-color images were generated for all the image types, where the image brightness is the average OCT intensity of the four images and the color (hue) of the image is one of GT, refined LIV, or non-refined LIV.
The details of the color image formation process can be found in Section 2.3 of Ref.\@ \cite{ElSadek2020BOE}.

To evaluate the refined NN model, we used 12 breast cancer spheroids.
Note that these spheroids are not the same as those used in the model refinement.
The spheroids were treated with DOX at concentrations of 0, 0.1, 1, and 10 \uM over treatment durations of 1, 3, and 6 days.
Namely, we used one spheroid for each condition.
Unlike the refinement process, the full-size image was fed into the trained NN model at once.
And hence, the full-size \enfaceh LIV image was obtained via a single inference operation.

\subsection{Demonstration of four-volume measurement}
\label{sec:4volAcqui}
Thus far, we have described methods based on datasets obtained by the 32-volume sequential acquisition protocol.
Namely, at each \enfaceh location, the NN-based LIV images were generated using four volumes extracted from a 32-volume sequence.
In real applications of the NN-based LIV image, we do not need to acquire 32 volumes, but four volumes are sufficient.
The implementation of the four-volume-acquisition protocol can reduce the data size.

To demonstrate the NN-based LIV image obtained with this four-volume acquisition, we designed another acquisition protocol that gives four volumes as depicted in Fig.\@ \ref{fig:acquiProto}(b).
These four volumes have the same inter-volume time separation (i.e., 2.15 s) as the four volumes extracted from the sequence obtained by the 32-volume acquisition.
Since one volume is acquired in 0.25 s, the total acquisition time is 7.45 s.

To experimentally demonstrate this protocol, we measured the same 12 spheroids used for the model evaluation described in Section \ref{sec:evaluationStudy}.
Namely, each of the 12 spheroids was measured twice using the 32- and four-volume acquisition protocols.

\section{Result}
\label{sec:result}

\subsection{Image-based model evaluation}
\label{sec:modelEvaluation}
\begin{figure}
	\centering\includegraphics[width=14cm]{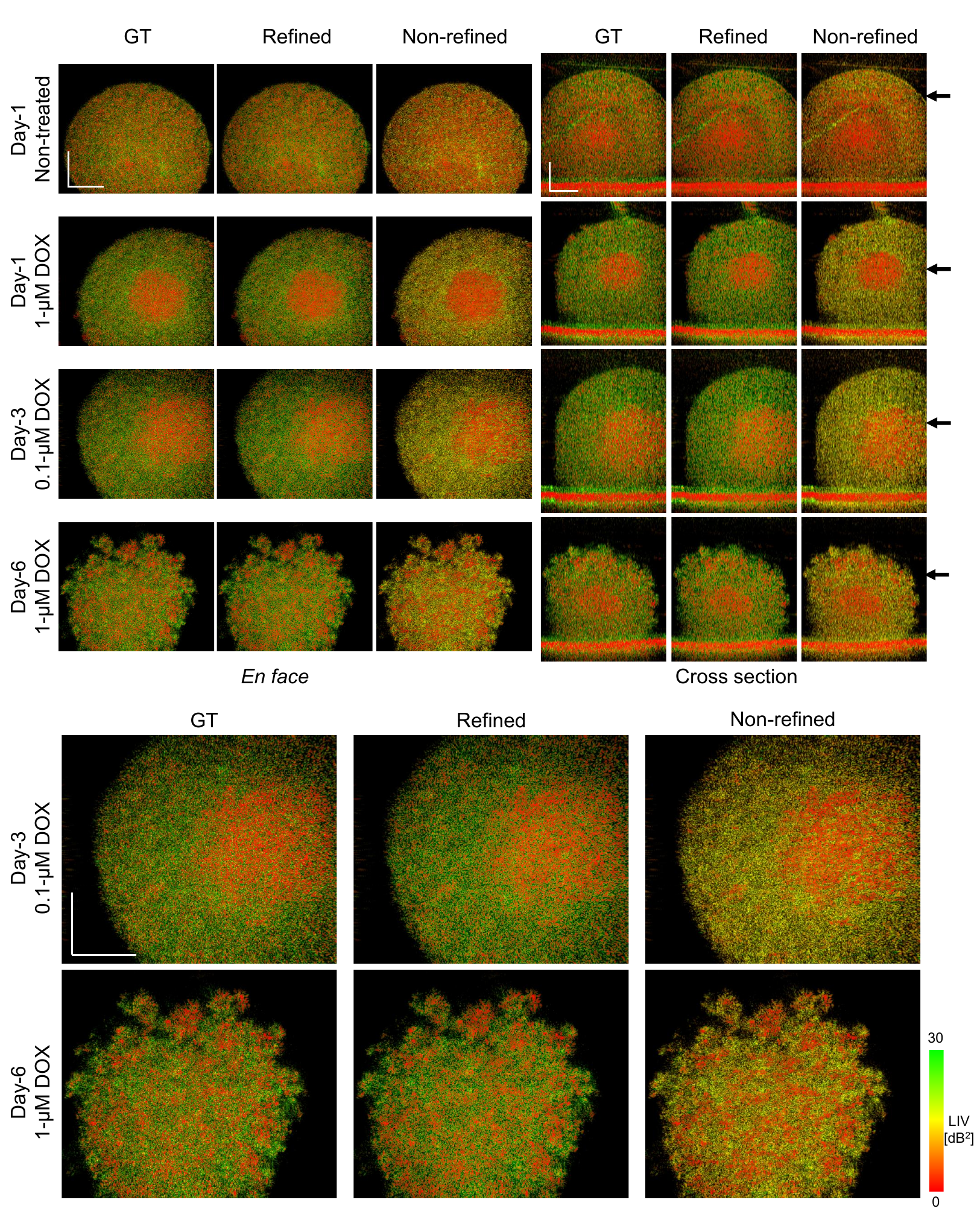}
	\caption{%
		Comparison of the GT, refined LIV, and non-refined LIV images of human breast cancer spheroids.
		The left and right halves of the figure show the \enfaceh and cross-sectional images, respectively.
		The depth positions of the \enfaceh images are indicated by black arrows on the right side of the cross-sectional images.
		The last two rows are magnified views of representative cases.
		The scale bar represents 100 \um.
		The output of the non-refined NN model exhibits lower LIV values than the GT and artifactual horizontal elongation patterns.
	}
	\label{fig:modelEvaluation}
\end{figure}
Figure \ref{fig:modelEvaluation} shows the \enfaceh and cross-sectional GT, refined LIV, and non-refined LIV images of four representative spheroids: non-treated spheroid, 1-\uM DOX-treated spheroid for 1 day, a 0.1-\uM DOX-treated spheroid for 3 days, and a 1-\uM DOX-treated spheroid for 6 days.

In comparison to the GT, the non-refined LIV has lower LIV values than the GT in the entire spheroid region.
However, the refined LIV shows quite similar values with the GT.
The detailed investigation of the output value differences is given in Section \ref{sec:optimalRefine}.

The low output values in the non-refined NN model can be explained by the difference of lateral resolutions between the two OCT systems and the resolution dependency of LIV \cite{Fujimura2025BOE}.
Specifically, point-scanning OCT used to train the basement NN model has resolutions of 19 \um (lateral) and 14 \um (axial), while FF-SS-OCM has those of 1.4 \um (lateral) and 4.7 \um (axial) as described in Sections \ref{sec:basementModel} and \ref{sec:DOCTacqui}.
Our previous study found that higher resolution leads to higher LIV values\cite{Fujimura2025BOE}. 
In fact, conventional (non-NN) LIV images of spheroids obtained by point-scanning OCT typically have values in the range of 0 to 10 \dbsq \cite{ElSadek2020BOE}, while LIV images of spheroids obtained by FF-SS-OCM have values in the range of 0 to 30 \dbsq.
Since the basement NN model was trained with point-scanning LIV images that have lower values than the LIV of FF-SS-OCM, the naive application of FF-SS-OCM data to the basement model can cause the lower output values than the GT.

In addition to the low output values, fine texture patterns in the non-refined LIV image differ from those of the GT.
Namely, the texture patterns in the non-refined image are elongated along the horizontal \enfaceh direction (that corresponds to the fast-scan direction of the point-scanning OCT).
These appearances are more distinct in the magnified images of day-3 0.1-\uM and day-6 1-\uM DOX treated spheroids (the last two rows in Fig.\@ \ref{fig:modelEvaluation}).
On the other hand, the LIV images of the refined NN model do not show this elongation artifact.

The elongated patterns may be caused by the different aspect ratio of pixels between the point-scanning OCT used for the basement-NN-model training and the FF-SS-OCM.
As described in Sections \ref{sec:basementModel} and \ref{sec:DOCTacqui}, the lateral pixel separations of the point-scanning OCT are non-isotropic as 1.95 \um (fast scan, horizontal in \enface) $\times$ 7.81 \um (slow scan, vertical), while those of the FF-SS-OCM is isotropic as 0.48 \um $\times$ 0.48 \um.
The four-times larger pixel separation of the point-scanning OCT along the horizontal direction can lead to the elongation pattern  when isotropic FF-SS-OCM data is input into the basement NN model.

\subsection{Demonstration of four-volume measurement}
\label{sec:resultFourVol}
\begin{figure}
	\centering\includegraphics[width=14cm]{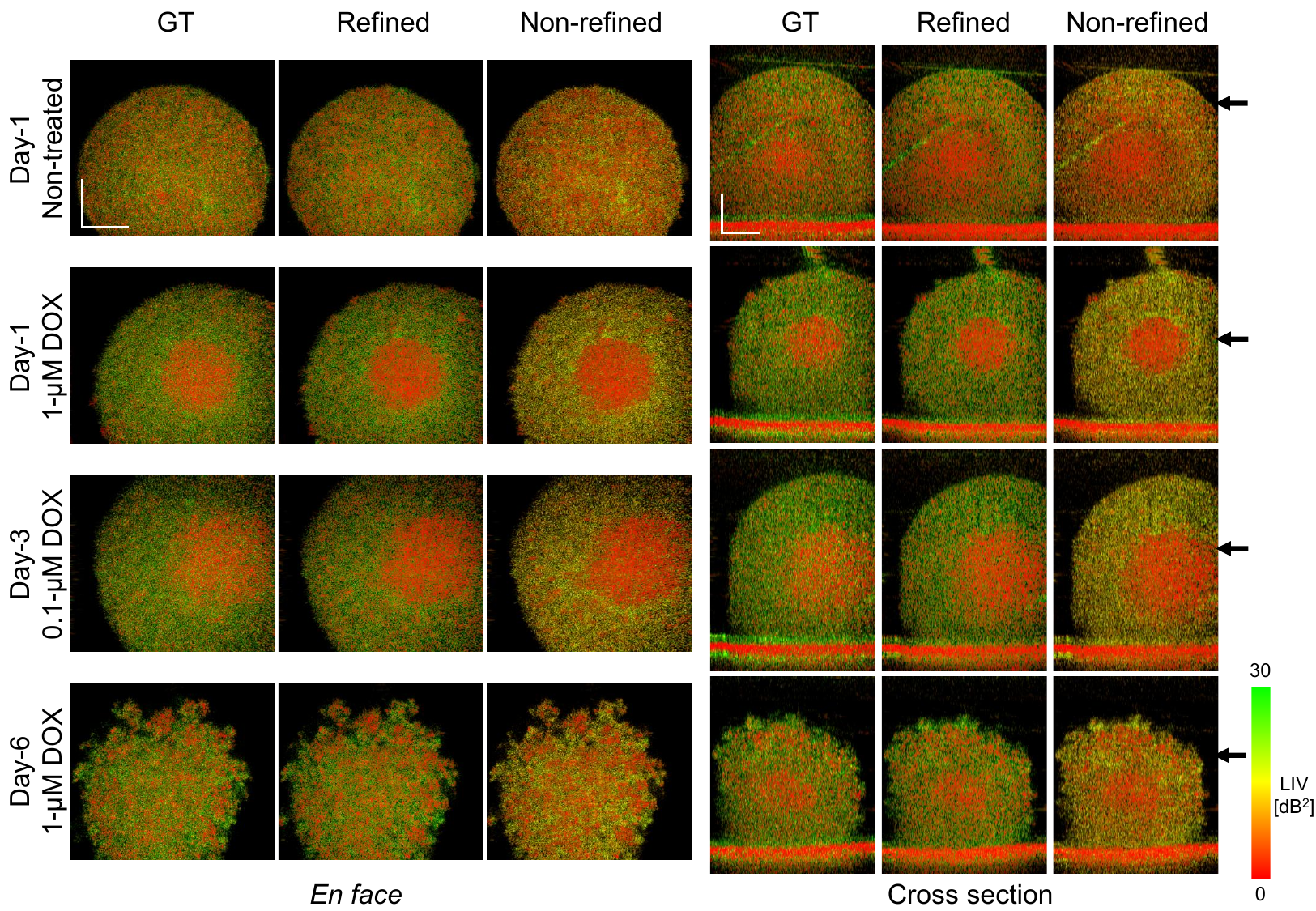}
	\caption{
		Comparison of the GT, refined LIV, and non-refined LIV images.
		Note that the GT images are generated from 32-volume sequence and are identical to those in Fig.\@ \ref{fig:modelEvaluation}, while the refined and non-refined images were generated from the dataset obtained by the four-volume acquisition protocol.
		The left and right halves of the figure show the \enfaceh and cross-sectional images, respectively.
		The depth positions of the \enfaceh images are indicated by black arrows on the right side of the cross-sectional images.
		The scale bar represents 100 \um.
		The refined images are similar to the GT.
		}
	\label{fig:4volumeMeasurement}
\end{figure}
Figure \ref{fig:4volumeMeasurement} demonstrates the \enfaceh and cross-sectional four-volume-acquisition-based LIV images of the same spheroids shown in Fig.\@ \ref{fig:modelEvaluation}, where GT, refined LIV, and non-refined LIV images are compared.
Note that the GT images in Fig.\@ \ref{fig:4volumeMeasurement} are identical to those in Fig.\@ \ref{fig:modelEvaluation}, namely, were computed from the 32-volume sequence.
On the other hand, the refined and non-refined LIV images are generated from the four-volume sequence acquired by the four-volume acquisition protocol described in Section \ref{sec:4volAcqui}.
The non-refined NN model again exhibits the same problems described in Section \ref{sec:modelEvaluation}, whereas the refined NN model resolves those problems and shows similar image appearances to the GT.

In addition to the image-based improvements, the implementation of the four-volume measurement significantly reduced data size, data streaming time, and data processing time.
The data size was around 42 GB for 32 complex OCT volumes, while it was reduced to only around 5.3 GB for the four volumes used for NN-based LIV generation.
Due to this data size reduction, the data transfer time from the camera to the PC was reduced from 7.15 min to only around 55 s.
Note that the computation of data transfer time was performed using a camera application made by the camera manufacture (Fastcam mini-AX Viewer PFV4, Photron).
The data processing time, including OCT reconstruction, computational refocusing, and DOCT computation, was also reduced from around 4 hours to 30 min.

\subsection{Loss function optimization for model refinement}
\label{sec:optimalRefine}
Thus far, the demonstration of the refined-NN-based LIV generation and four-volume acquisition has been discussed based on the refined NN where the loss was calculated by wMAE, which weighted LIV values of 30 \dbsq or more by a factor of 3 (see Section \ref{sec:detailedImplementaion}).
To investigate whether this loss function is reasonable or not, we performed several cases of model refinements with different loss functions.

\subsubsection{Study protocol}
In this study, three types of loss functions were compared: MAE, wMAE, and mean squared logarithmic error (MSLE).
MSLE is known to be suitable for data with a wide range of values, such as LIV images generated from FF-SS-OCM data, and is defined as follows
\begin{equation}
	\text{MSLE} = \frac{1}{N}\sum_{x, y, b} \left[ \log(\text{LIV}(x,y;b)+1) - \log(\text{LIV}'(x,y;b)+1) \right]^2,
\end{equation}
where $N$ is the sum of $x$, $y$, and $b$.
Note that we used two types of wMAEs.
One is identical to the wMAE used for the studies in Sections \ref{sec:modelEvaluation} and \ref{sec:resultFourVol}, which uses a weighting factor of 3.
The other is a wMAE that weights LIV values of 30 \dbsq or more by a factor of 2, instead of 3.
Hereafter, the wMAEs with weights of 2 and 3 are denoted as w2MAE and w3MAE, respectively.

To evaluate all the four types of refined NN models, the same spheroid datasets used for model evaluation described in Section \ref{sec:evaluationStudy} were input into each refined NN model to obtain output images.
From each output image, regions of interest (ROIs) were selected in the spheroid periphery and core as shown in the GT images in Fig.\@ \ref{fig:trainOptimiz}.
The size of the ROIs was around 100 $\times$ 100 pixels. 
Then, mean LIV values in the ROIs of the periphery and core for each of the 12 spheroid datasets were computed, and all the mean values were averaged in the periphery and core over the 12 spheroids.
In addition, the root mean squared (RMS) differences between the ROI averages of the GT and each of the refined and non-refined were computed again over the 12 spheroids.

\subsubsection{Observational and quantitative comparison of refined NN model performance}
\begin{figure}
	\centering\includegraphics[width=13cm]{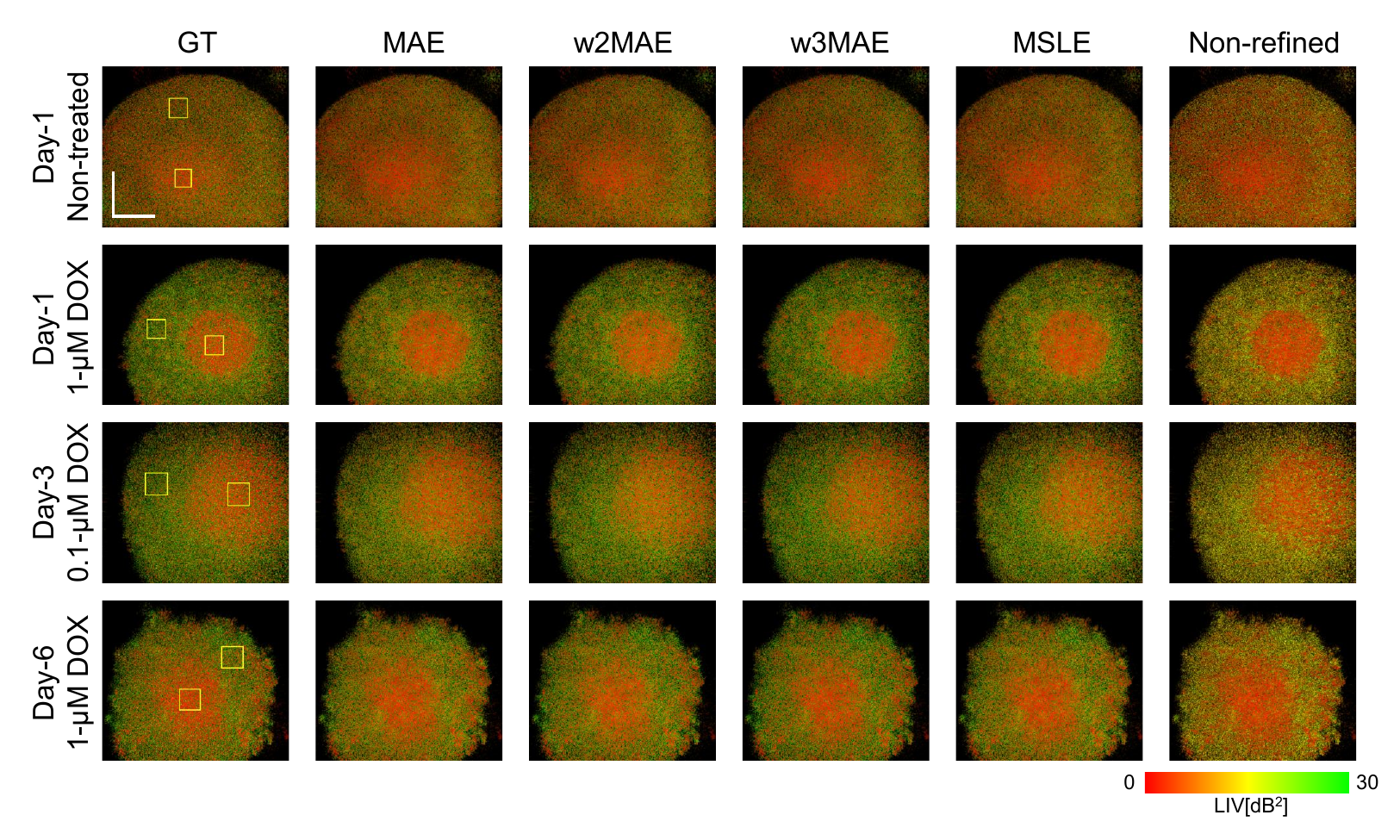}
	\caption{
		Observational comparison of output images generated from several types of refined NN models.
		In the ground truth images, ROIs are indicated by yellow squares.
		The scale bar represents 100 \um.
		In the eye observation, all the refined NN models generated similar images to the GT.
	}
	\label{fig:trainOptimiz}
\end{figure}
\begin{figure}
	\centering\includegraphics[width=14cm]{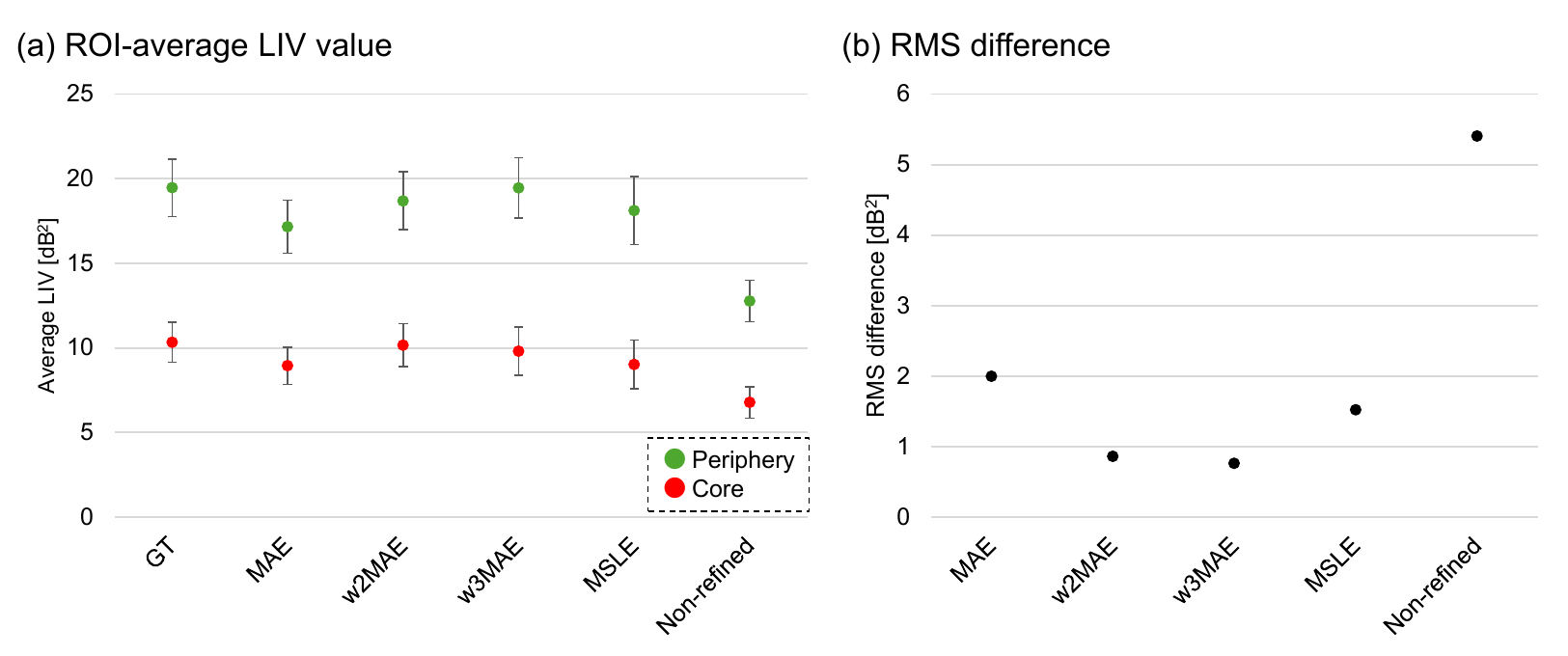}
	\caption{%
		Quantitative comparison of LIV values generated by several types of refined NN models.
		(a) ROI average LIV values at periphery and core ROIs.
		The circles and bars indicate the mean and the standard deviation of ROI average LIV values among the 12 spheroid samples, respectively. 
		The w2MAE and w3MAE models show superior performance to the other refined NN models.
		(b) RMS difference of mean of ROI average LIV values from the GT.
		Again, the w2MAE and w3MAE models show superior performance.
		Since the w3MAE model shows the smallest value, it is reasonable to be used for the refinement process.
	}
	\label{fig:quantiEval}
\end{figure}

Figure \ref{fig:trainOptimiz} shows the output images of the four refined NN models and the non-refined NN model of the four representative samples: non-treated spheroid, 1-\uM DOX-treated spheroids for 1 day, a 0.1-\uM DOX-treated spheroid for 3 day, and a 1-\uM DOX-treated spheroid for 6 day.
The difference between the GT and each refined LIV seems to be not significant by the eye observation.
However, in terms of the output values, some models exhibit significant differences from the GT as shown in Fig.\@ \ref{fig:quantiEval}.

Figure \ref{fig:quantiEval}(a) shows ROI average LIV values computed from the GT and refined LIV images.
The circles and the error bars indicate the mean and the standard deviation of the ROI averages among the 12 spheroid samples, respectively.
In the periphery region (i.e., periphery ROI), the w3MAE model outputs the closest values to the GT among all the refined NN models.
On the other hand, in the core region, the w2MAE model shows the best performance, and the w3MAE model is the second best.
Figure \ref{fig:quantiEval}(b) shows the RMS difference from the GT.
The w2MAE and w3MAE models show better performance than the other refined NN models.
Among them, the w3MAE model shows the lowest RMS difference.
Based on these quantitative comparisons, we concluded that the w3MAE model is a reasonable option and have used it as the primary choice.

\section{Discussion}
\subsection{Application to other type of spheroids}
\label{sec:applicationToHT-29}
\begin{figure}
	\centering\includegraphics[width=14cm]{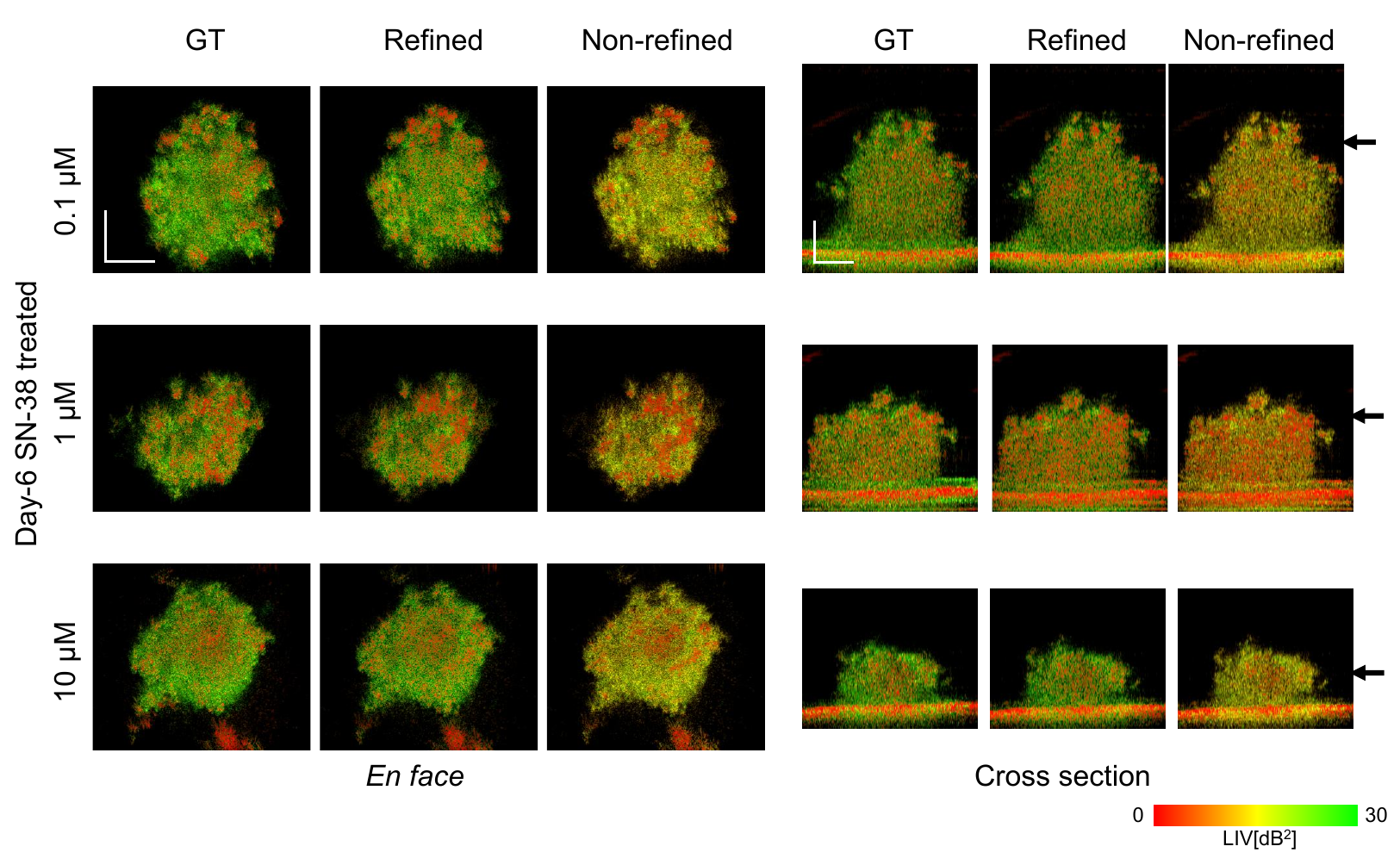}
	\caption{Application of the refined NN model to colon cancer spheroids treated with SN-38 at concentrations of 0.1, 1, and 10 \uM for 6 days.
	The left and right halves of the figure show the \enfaceh and cross-sectional images, respectively.
	The depth positions of the \enfaceh images are indicated by black arrows on the right side of the cross-sectional images.
	The scale bar represents 100 \um.
	The non-refined NN model generates some elongated low-LIV regions and outputs lower LIV values than ground truth, while the refined NN model resolves these problems and generates similar image appearances to the GT.}
	\label{fig:4volBasedHT-29spheroid}
\end{figure}
In the main study, the refined NN model was trained and evaluated with human breast cancer spheroid data.
Here, we investigate the applicability of the refined NN model to another type of spheroid using colon cancer (HT-29 cell line) spheroids.
The colon cancer spheroids were treated with anti-cancer drug (SN-38) at concentrations of 0.1, 1, and 10 \uM for 6 days and measured by FF-SS-OCM with the 32-volume and four-volume acquisition protocols described in Section \ref{sec:4volAcqui}.
The GT was computed from the sequential 32 volumes.
Refined LIV images were obtained by inputting the four volumes obtained by the four-volume-acquisition protocol to the refined NN model.
Similarly, non-refined LIV images were obtained from the four volumes.

Figure \ref{fig:4volBasedHT-29spheroid} shows the GT, refined LIV, and non-refined LIV of the colon cancer spheroids.
The drug-treated colon cancer spheroids show a different drug response compared to the breast cancer spheroids.
The GT images show that, unlike the breast cancer spheroids, the SN-38-treated colon cancer spheroids do not show a clear core and do not form a spherical shape.
Although this type of spheroid was not used in the refinement process, the refined LIV shows consistent image appearances with the GT, whereas the non-refined LIV shows lower LIV values than the GT.
These results suggest that the refined NN model has been generalized to the different types of spheroids to some degree.

\subsection{Limitations of current refined NN-based LIV generation method}
There are some limitations and open issues with the current refined NN-based DOCT generation method.
One limitation is that the current method is specific to a single type of sample, i.e., tumor spheroids.
Additionally, the current method is applicable to only a specific type of DOCT contrast, i.e., LIV.

Regarding the first issue, although the current refined NN model was found to be generalized to some degree (Section \ref{sec:applicationToHT-29}), it might worth to further generalize the model.
Additional training or refinements with other types of samples, such as other types of spheroids and also organoids may further generalize the model in the future.

For the second issue, it should be noted that some DOCT algorithms, such as amplitude-spectrum-based DOCT (AS-DOCT) \cite{Munter2020OL}, an inverse-power-law-based method \cite{Oldenburg2015Optica}, OCT correlation decay speed \cite{ElSadek2020BOE}, and Swiftness\cite{Morishita2025BOE}, exploit the time-order information of the image sequence, while LIV (i.e., the time variance) is not sensitive to the time-order of the signals. 
The extension of the NN-based DOCT to such time-order sensitive DOCT method may require a more sophisticated architecture and/or a well-designed data representation.
Recently, NN-based AS-DOCT was demonstrated using non-time-uniform data acquisition and a three-dimensional (space and time) convolutional network \cite{Liu2025BiOS}.
Similarly, NN-based Swiftness measurement was achieved by combining the non-time-uniform acquisition with a dual input of the image time sequence and the acquisition-time-dependency of the signal variance \cite{Liu2026BiOS}.
Although these methods are based on point-scanning OCT, they could be applicable to FF-SS-OCM with some modifications.

\section{Conclusion}
We have demonstrated an NN-based DOCT method for high-resolution FF-SS-OCM that generates LIV images from only 4 OCT volumes.
By refining the already established (basement) NN model using a small number of FF-SS-OCM dataset (only 20 samples), the refined NN model became applicable to FF-SS-OCM.
Additionally, the demonstration of the four-volume acquisition protocol significantly reduced data size, data transfer time, and processing time by a factor of eight.

Since FF-SS-OCM provides high-definition but large-data-size OCT volumes, the reductions of the data size, the data-transfer time, and the processing time are significant to translate FF-SS-OCM-based DOCT to the practical applications.

\begin{backmatter}
\bmsection{Funding}
Japan Science and Technology Agency (JPMJCR2105, JPMJSP2124); Japan Society for the Promotion of Science (21H01836, 22F22355, 22KF0058, 22K04962, 24KJ0510).

\bmsection{Disclosures}
Komeda, Tateno, Liu, Morishita, Wang, El-Sadek, Makita, Yasuno: Nidek(F), Sky Technology(F), Panasonic(F), Nikon(F), Santec(F), Kao Corp.(F), Topcon(F).
Furukawa, Matsusaka: None.
Tateno is currently employed by KOWA.

\bmsection{Data availability}
Data underlying the results presented in this paper are not publicly available at this time but may be obtained from the authors upon reasonable request.

\end{backmatter}

\bibliography{NNbasedDOCT}

\end{document}